\documentclass[twocol]{ametsocV6.1}

\usepackage{soul}
\title{Effects of high-frequency and balanced motions on Lagrangian pair dispersion at the ocean surface} 

\authors{Michael Maalouly\aff{a}, 
Apolline Dekens\aff{a}, 
Guillaume Lapeyre\aff{b}, 
Aur\'elien Luigi Serge Ponte\aff{c}, 
Stefano Berti\aff{a}\correspondingauthor{Stefano Berti, stefano.berti@polytech-lille.fr}
}

\affiliation{\aff{a}{Univ. Lille, ULR 7512 - Unit\'e de M\'ecanique de Lille Joseph Boussinesq (UML), F-59000 Lille, France}\\
\aff{b}{LMD/IPSL, CNRS, Ecole Normale Sup\'erieure, PSL Universit\'e, 75005 Paris, France}\\
\aff{c}{Laboratoire
d'Oc\'eanographie Physique et Spatiale, CNRS, IFREMER, IRD, IUEM, Universit\'e de Brest, Brest, France}
}

\abstract{We investigate the properties of relative dispersion of Lagrangian particles in a global-ocean simulation resolving both inertia-gravity waves (IGW) and meso and submesoscale (M/SM) turbulence. More specifically, we test if the dispersion laws depend on the shape of the Eulerian kinetic energy spectrum, as predicted from quasi-geostrophic turbulence theory.
To this end, we focus on two areas, 
in the Kuroshio Extension and in the Gulf Stream, for which the relative importance of IGW compared to M/SM vary in summer and winter.
In winter, Lagrangian statistical indicators return a picture in overall agreement with the shape of the kinetic energy spectrum. Conversely, in summer, when submesoscales are less energetic and higher-frequency internal waves gain importance, the expected relations between dispersion properties and spectra do not seem to hold. 
This apparent discrepancy is explained by decomposing the flow into nearly-balanced motions and internal gravity waves, and showing that the latter dominate the kinetic energy spectrum at small scales.
Our results are consistent with the hypothesis that high-frequency IGWs do not impact relative dispersion,  which is then controlled by the nearly-balanced, mainly rotational, flow component at larger scales. 
These results highlight that geostrophic velocities derived from wide-swath altimeters, such as SWOT,  may present limits when estimating surface dispersion, and that current measuring satellite missions may provide the complementary information to do so.
}

\begin{document}

\maketitle

%%%%%%%%%%%%%%%%%%%%%%%%%%%%%%%%%%%%%%%%%%%%%%%%%%%%%%%%%%%%%%%%%%%%%
% MAIN BODY OF PAPER
%%%%%%%%%%%%%%%%%%%%%%%%%%%%%%%%%%%%%%%%%%%%%%%%%%%%%%%%%%%%%%%%%%%%%

%%%%%%%%%%%%%%%%%%%%%%%%%%%%%%%%%%%%%%%%%%%%%%
\section{Introduction}\label{sec:intro}

Ocean flows at lengthscales smaller than few hundreds of kilometers are composed of a rich variety of dynamical structures, e.g. fronts, eddies and internal gravity waves (IGW). On one hand, fronts and eddies constitute the so-called meso and submesoscales (M/SM), which evolve over timescales of days to weeks. On the other, IGWs are associated with more rapid (of the order of hours) temporal dynamics, which tend to interact with, and dampen, lower-frequency balanced motions~\citep{barkan2017}. 
Understanding these interactions is important, for instance for the interpretation and exploitation of new, high-resolution satellite-altimetry data \citep{uchida2024} or the characterization of material transport at fine scales \citep{holmescerfon_2013,hernandezduenas_2021}.

Although high-frequency motions, such as internal tides and gravity waves, are often considered to weakly contribute to the transport of tracers~\citep[see, e.g.,][]{Beron2016}, their effect on the dispersion of Lagrangian drifters remains poorly explored~\citep{lumpkin2017advances}, and the results do not seem completely conclusive. Relying on high-resolution numerical simulations in the south Atlantic ocean, it was argued that high-frequency motions considerably increase Lagrangian diffusivity, particularly at small scales~\citep{sinha19}. 
However,~\citet{Wangetal_2018}, using a non-hydrostatic numerical model representing both an upper mixed layer and internal waves, showed that while high-frequency motions may have an effect on pair dispersion rates, the details of this effect depend on the specific features of the M/SM dynamics. 

Beyond their interest for material transport, Lagrangian studies also reveal useful to characterize the submesoscales, as shown in many regions of the world ocean. 
The link between Lagrangian measurements and statistical properties, as those quantified by the 
kinetic energy spectrum of the underlying flow, is then established through different bridging relations, obtained dimensionally in the framework of classical fluid-turbulence theory. The utility of this approach for quasi-geostrophic (QG) meso and submesoscale dynamics is well documented~\citep{lacasce2008, berti2011lagrangian,Poje_etal_2014,CLPSZ2017,FBPL2017}. 
Assessing whether high-frequency motions affect particle dispersion regimes, and their possible impact on the validity of these bridging relations, thus remains a question of prime scientific interest.

If a possible limitation of Lagrangian data is their moderately sparse coverage, a global view of ocean-surface currents can be achieved through satellite-altimetry measurements. Conventional instruments, however, were limited in spatial resolution to $O(100)$~km~\citep{morrow2023ocean}, which has not permitted, so far, the observation of structures in the submesoscale range, or even in the lower end of the mesoscale one. 
New-generation, wide-swath altimetry is pushing this limit to much smaller scales. Indeed, the Surface Water and Ocean Topography (SWOT) mission has recently started to provide sea surface height (SSH) data at an unprecedented resolution of $(5-10)$~km ~\citep{fu2024surface}. 
While this represents a major advancement in our ability to access the fine-scale range, the proper exploitation of these data also raises several important challenges. For instance, oceanic currents are retrieved from SSH assuming geostrophic balance. However the latter is not granted to hold at the smallest resolved scales, where ageostrophic and high-frequency motions may be expected to have a non-negligible dynamical signature~\citep{yu2021geostrophy}. 
Determining with what accuracy (in terms of spatial scales) the velocity fields computed from SSH represent real surface currents, and their turbulent properties, then seems crucial. 
An interesting approach to address this point is to examine Lagrangian statistics, which reflect the temporal evolution of fluid parcels in the
flow and hence sample processes acting on different timescales. 
This can be done, for instance, by comparing Lagrangian statistics from synthetic drifters advected by SWOT-derived velocities and real drifters~\citep{tranchant2025swot}. 
Another avenue of efforts, which is the one undertaken here, is to resort to high-resolution numerical simulations and to compare Lagrangian dispersion properties with their predictions from QG turbulence theory. In this case, the availability of the velocity field at high spatial and temporal resolution is expected to ease correlating Lagrangian diagnostics and Eulerian flow properties and, in the end, to disentangle contributions from the different physical processes at play. 

In this work, we use high-resolution velocity fields from the MITgcm LLC4320 simulation~\citep{forget2015ecco}, resolving submesoscales and accounting for IGWs, to advect Lagrangian tracer particles. We then characterize relative-dispersion statistics using different types of indicators, namely computed either at fixed time or at fixed lengthscale. 
More specifically, we aim to assess whether and how high-frequency motions impact the behavior of Lagrangian diagnostics, particularly testing the relation of the latter 
with the spectral kinetic energy of the Eulerian flow.
We focus on the Kuroshio Extension region and examine the seasonal dependence of the results. As winter and summer lead to distinct features in terms of meso and submesoscale energetics, this will allow us to explore the sensitivity of the difference in intensity of M/SM motions compared to IGWs.  
In order to test the generality of our main results, we then perform the same analysis also in another energetic region, close to the Gulf Stream.

This study extends previous ones~\citep{maaloulyetal2023,maaloulyetal2024},  conducted in the framework of the idealized SQG$^{+1}$ model, a quasi-geostrophic model including next-order corrections in the Rossby number~\citep{lapeyre2017,HSM2002}.  
Those studies showed that including the ageostrophic flow component into particle advection has quite marginal effects on relative dispersion over long times~\citep{maaloulyetal2023,maaloulyetal2024}. However, by construction, the SQG$^{+1}$ model only accounts for weak deviations from geostrophic balance and, therefore, does not include internal waves, which motivates the present investigation. 

This article is organized as follows. Section~\ref{sec:llc_model} describes  LLC4320 simulation and the setup of the Lagrangian-advection numerical experiments. 
Section~\ref{sec:llc_eulerian} provides a characterization of the flow properties from Eulerian diagnostics in Kuroshio Extension. 
In Sec.~\ref{sec:llc_lagrangian} we examine the related Lagrangian pair-dispersion statistics. We then interpret these results through a decomposition of fluid motions into 
their IGW and M/SM components, relying on the computation of frequency-wavenumber energy spectra, in Sec.~\ref{sec:llc_igw_bm}. 
A discussion on the comparison with the results in the Gulf Stream region is provided in Sec.~\ref{sec:llc_gs}  and conclusions are drawn in Sec.~\ref{sec:concl}.

%%%%%%%%%%%%%%%%%%%%%%%%%%%%%%%%%%%%%%%%%%%%%%
\section{Numerical simulations}\label{sec:llc_model}

To explore the impact of high-frequency motions and submesoscales on Lagrangian dispersion, we use data from the global-ocean LLC4320 simulation~\citep{forget2015ecco} to simulate trajectories of synthetic particles. 
LLC4320 was performed using MITgcm~\citep{marshall1997finite} with a horizontal spatial resolution of $1/48^{\circ}$, corresponding $\approx 0.75$~{km} in polar regions to $\approx 2.2$~{km} in equatorial ones. This resolution allows to resolve mesoscale dynamics and, to good extent, submesoscale ones. The model is tidally forced at different frequencies and was shown to reproduce diurnal and semidiurnal tidal variances with moderate biases compared to surface drifters \citep{yu2019surface, arbic2022near,caspar2025combining}. 
The output fields are available at hourly time intervals for a $1$-year period spanning from September $13$, $2011$ to November $15$, $2012$. The model capabilities to realistically account for the above mentioned physical processes were extensively discussed in previous studies~\citep[see, e.g.,][]{Torresetal_2018,Torresetal_2022,yu2019surface,yu2021geostrophy}. 
Here we focus on the dynamics of Lagrangian tracer particles at the ocean surface. Particle advection is performed offline by means of the Python OceanParcels package~\citep{lange2017parcels,zhang2024lagrangian}, using the surface velocity fields extracted from LLC4320 simulation. 
The Lagrangian evolution equations are integrated
using a fourth-order Runge-Kutta method and TRACMASS in space of the velocity
field at particle positions~\citep{doos2017evaluation,delandmeter2019parcels}. 
 
In the following, we will examine two regions of the ocean (Kuroshio Extension and Gulf Stream). For each region,
inside a square of side $\approx 500$~km (as in Fig.~\ref{fig:f1}a,c), 
$N=3600$ particles are initially uniformly distributed
in triplets, each arranged in an equilateral triangle inscribed within a circle of radius $1$~{km}. After their seeding, particles are tracked in time for a 30-day period during both February and August $2012$,  with hourly temporal resolution. For the statistical analysis of the relative dispersion process we consider only original pairs, meaning having a prescribed separation distance $R_0$ at the seeding time.

%%%%%%%%%%%%%%%%%%%%%%%%%%%%%%%%%%%%%%%%%%%%%%
\section{Eulerian flow properties of the Kuroshio Extension region}\label{sec:llc_eulerian} 

\begin{figure*}
    \centering
    \includegraphics[width=0.9\linewidth]{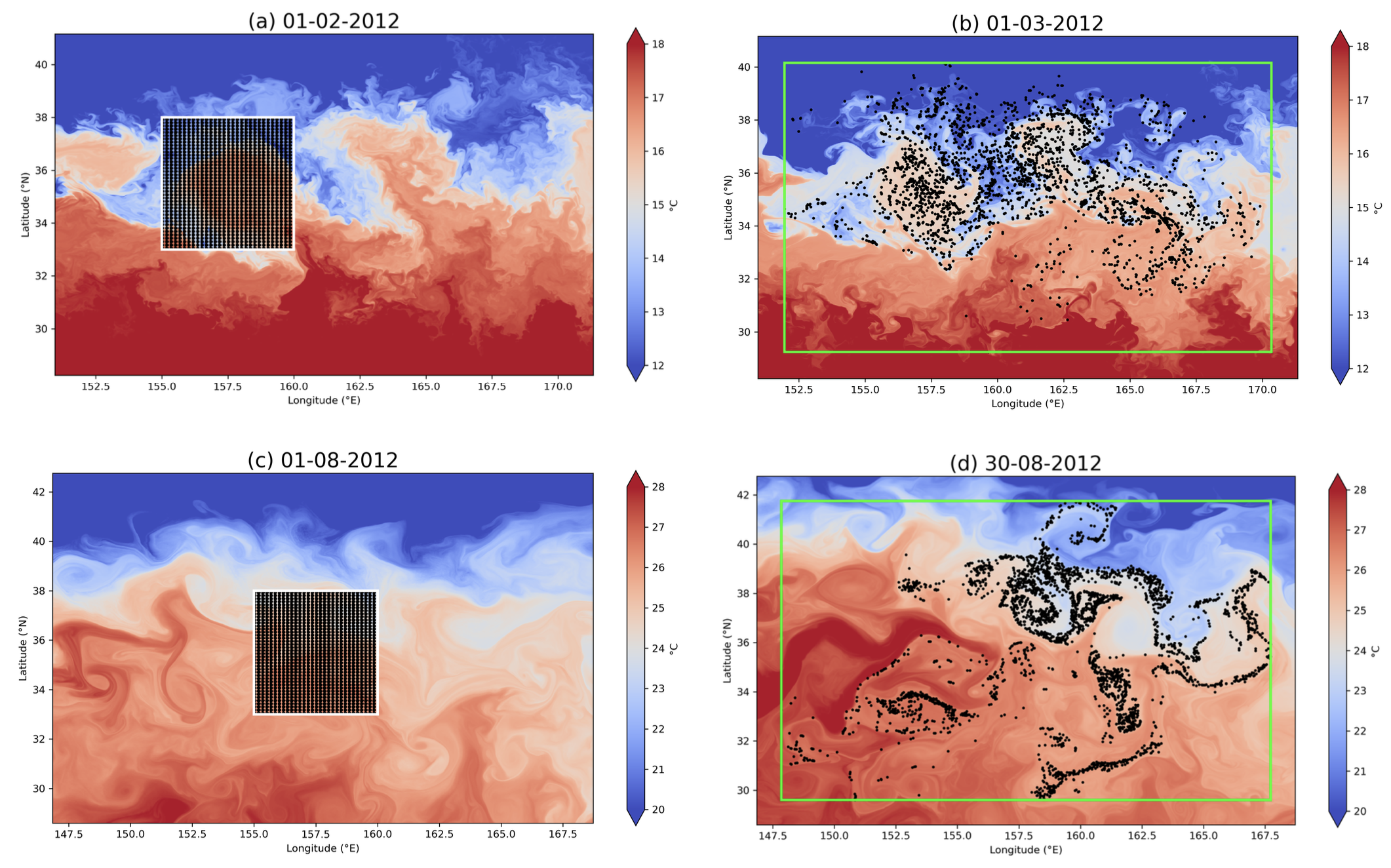}
    \caption{Snapshots of the SST field in the Kuroshio Extension region in February (top line)  and in August (bottom line) at the beginning (a, c) and at the end (b, d) of the $30$-day long Lagrangian experiments. The corresponding particle distributions are shown with black dots. The green rectangles in (b, d) indicate the largest area covered by particles on the latest day of the month. 
    }
    \label{fig:f1}
\end{figure*}

\begin{figure}
    \centering
    \includegraphics[width=\linewidth]{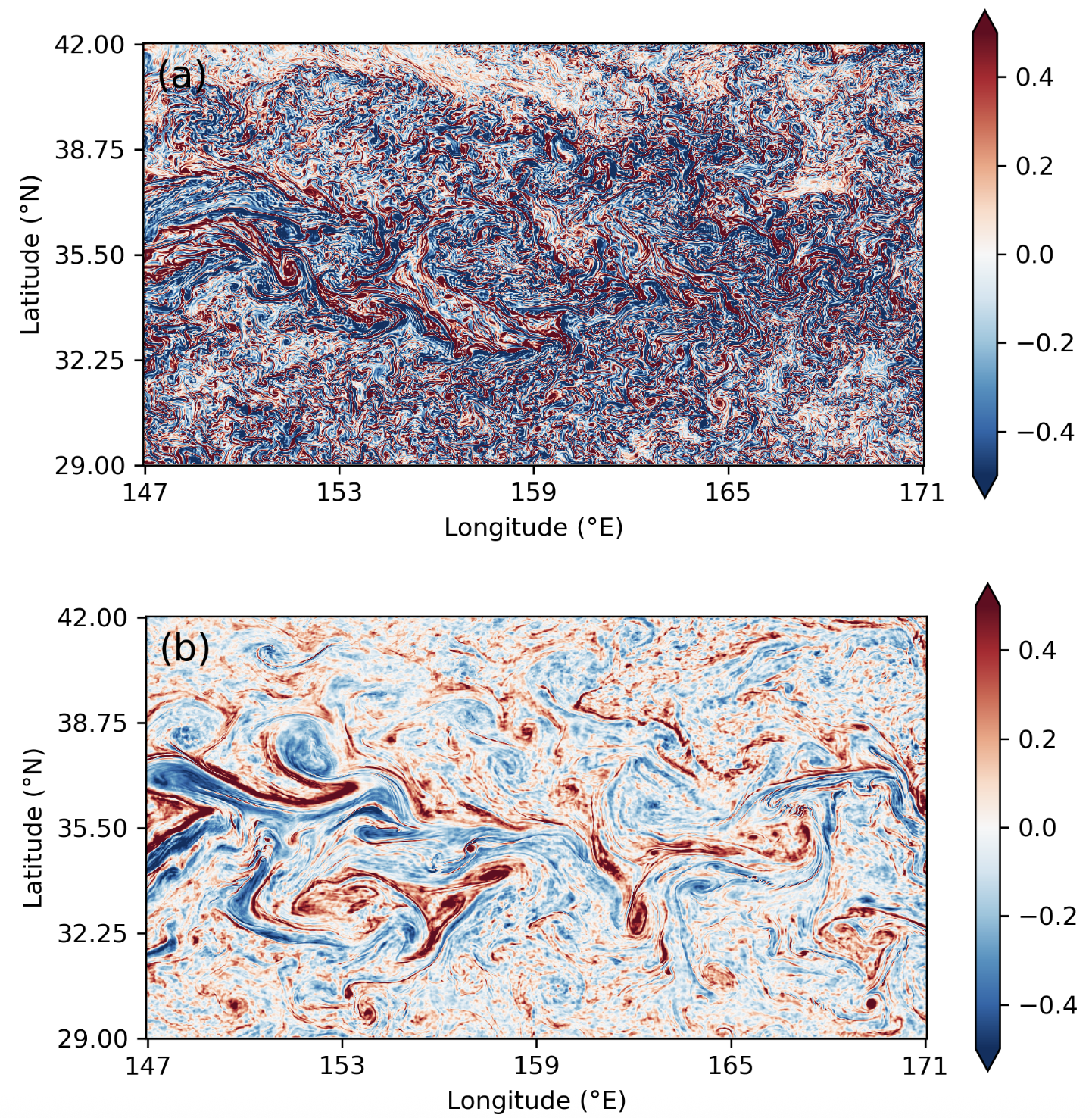}
    \caption{Snapshots of relative vorticity, normalized by the Coriolis parameter, $\zeta/f$ in the Kuroshio Extension region, for February 15, 2012 (a) and August 15, 2012 (b).}
    \label{fig:f2}
\end{figure}

We start our analysis by presenting the region we are focusing on.
Figure~\ref{fig:f1} shows sea surface temperature (SST) in both February (top row) and August (bottom row), at the beginning (left) and at the end (right) of the particle advection experiments. 
In both seasons, the region encompasses the Kuroshio current, as seen through its associated large-scale SST gradient with warm (cold) waters on the equatorward (poleward) side of the jet. 
In February, large meanders of the SST front indicate
the presence of mesoscale structures, with a typical size of $150$ to $400$~km (Fig.~\ref{fig:f1}a,b). 
In addition,  a wealth a smaller eddies of $O(10)$~km size, due to submesoscale instabilities, can also be  seen along the large-scale SST front. On the contrary, in August (Fig.~\ref{fig:f1}c,d) the latter fine scales seem to fade out. These observations are confirmed by inspection of relative-vorticity snapshots, shown in Fig.~\ref{fig:f2} 
at mid February and August (i.e. half the total Lagrangian integration time). 
While in winter a dense population of submesoscale eddies and filaments is clearly visible, to the point that larger scales are hardly detectable, in summer vorticity is mainly concentrated at mesoscales and has a smoother, much more filamentary structure. Note, too, the weaker SST gradients in August compared to February as well as smaller values of relative vorticity. 
Such seasonality is consistent with past numerical~\citep{sasaki2014impact} and observational studies~\citep{callies2015seasonality}.
%%%%%%%%%%%%%%%%%
% FIG. 1
%%%%%%%%%%%%%%%%%
%%%%%%%%%%%%%%%%%
% FIG. 2
%%%%%%%%%%%%%%%%%

Since we aim at understanding how the behaviors of Lagrangian-dispersion indicators depend on the Eulerian flow properties, it is important to properly select the spatial domain over which the latter are computed. Considering that particles spread in time and distribute over a wider region than the one in which they were released (see Fig.~\ref{fig:f1}), for each month we decided to choose an area including all the 3600 particles at the end of the Lagrangian-tracking experiment (green rectangles in Fig.~\ref{fig:f1}b and Fig.~\ref{fig:f1}d). This ensures that Eulerian statistics reflect the properties of the velocity field sampled by Lagrangian tracers. 

\begin{figure}
    \centering
    \includegraphics[width=0.9\linewidth]{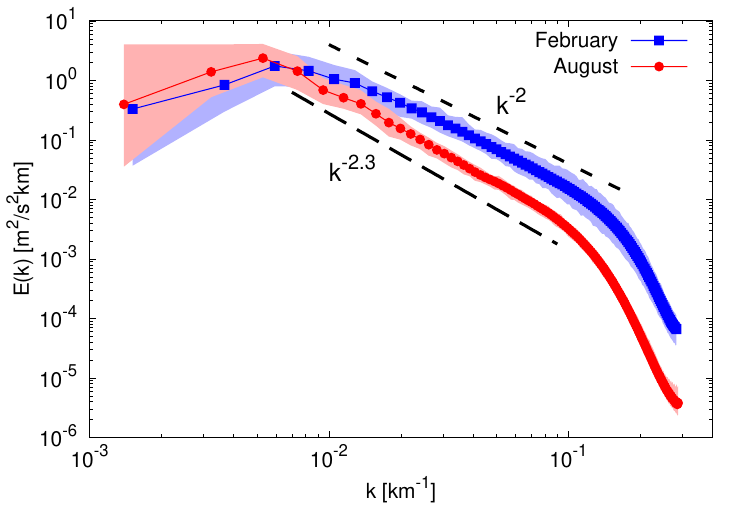}
    \caption{Wavenumber spectra of horizontal kinetic energy,  averaged over February (blue squares) and August (red dots), in the Kuroshio Extension region. For each month, the shaded areas represent the temporal variability of the spectrum. The reference lines $k^{-2}$ and $k^{-2.3}$ are also shown for comparison. 
    } 
    \label{fig:f3}
\end{figure}

The wavenumber spectra of horizontal kinetic energy, averaged in time over February and August are presented in Fig.~\ref{fig:f3}. They confirm that the flow in February is more energetic than in August, particularly at scales smaller than $100$~km. 
The winter kinetic energy spectrum scales approximately as $E(k)\sim k^{-2}$, as often observed in the presence of energetic submesoscales~\citep{Klein_etal_2008,Capet_etal2008}, 
over slightly more than a decade of wavenumbers.
Note, however, that due to the non-negligible uncertainties on $E(k)$, particularly at small scales, the spectral slope $\beta$ from a fit [for $k$ between $O(10)$~km and $O(100)$~km] varies in the range $5/3 \lesssim \beta \lesssim 2.4$, depending on the specific extension of the fitting range.   
The summer spectrum is characterized by smaller uncertainties (except at the largest scales), and its scaling behavior is close to $k^{-2.3}$ over a wavenumber range of comparable width. 
%%%%%%%%%%%%%%%%%
% FIG. 3
%%%%%%%%%%%%%%%%%

%%%%%%%%%%%%%%%%%%%%%%%%%%%%%%%%%%%%%%%%%%%%%%
\section{Lagrangian pair-dispersion statistics in Kuroshio Extension}\label{sec:llc_lagrangian}

After having described the main features of the Eulerian flow, we present in this section the results about Lagrangian pair-dispersion statistics. We recall that we consider only original pairs, with an initial separation distance $R_0 \approx 3.48$~km. Distances between particles at different times are computed on the sphere using Haversine formula. Uncertainties on the considered indicators are estimated as the $95$\% confidence interval of the bootstrapped mean of $1000$ samples.

A natural approach to analyze pair-separation processes is to measure the mean-square relative displacement between two particles as a function of time, i.e. relative dispersion 
%%%%%%%%%%%%%%%%%%%
\begin{equation}
\langle R^2(t)\rangle = \langle |\mathbf{x}_i(t)-\mathbf{x}_j(t)|^2 \rangle.
\label{eq:reldisp}
\end{equation}
%%%%%%%%%%%%%%%%%%%
In the above expression, $i=1,...,N$ labels a given particle among the $N$ considered ones, whose position evolves according to $\dot{\mathbf{x}}_i=\mathbf{u}(\mathbf{x}_i(t),t)$, with $\mathbf{u}=(u,v)$ the horizontal surface velocity. The angular brackets indicate an average over all $i$ and all corresponding particles $j$ with initial separation $|\mathbf{x}_i(0)-\mathbf{x}_j(0)|=R_0$, so that $\langle R^2(0)\rangle=R_0^2$.

We first recall the expected behavior of $\langle R^2(t) \rangle$ obtained from  dimensional arguments, for homogeneous isotropic incompressible two-dimensional turbulence. 
As extensively documented~\citep[see, e.g.,][]{babiano_etal_1990,FBPL2017} these expectations may be difficult to observe for different reasons, such as a finite inertial range of the energy and enstrophy cascades, or the sensitivity of $\langle R^2(t) \rangle$ to the distance of the initial pair separation.
At short enough times, relative dispersion is expected to grow in a ballistic way, $\langle R^2(t) \rangle \simeq R_0^2 + Z\,R_0^2\, t^2$~\citep{batchelor1950,babiano_etal_1990}. Here $Z= \langle \zeta^2/2 \rangle_x$ is relative enstrophy, $\langle ... \rangle_x$ denotes a spatial average, and vorticity is related to the horizontal flow by $\zeta=\partial_x v-\partial_y u$. Later in time, when the pair separation distance is intermediate between the smallest and the largest eddy sizes, the temporal growth of $\langle R^2(t) \rangle$ can be dimensionally linked to the shape of the kinetic energy spectrum $E(k)$. 
Assuming a power-law scaling $E(k) \sim k^{-\beta}$, if the spectrum is sufficiently steep ($\beta>3$) relative dispersion should grow exponentially in time, with a rate proportional to $Z^{1/2}$. 
Such fast decay of kinetic energy with wavenumber,  typical of weakly-energetic submesoscales, implies that strain is localized at large scale and, hence, that 
the pair-separation process is controlled by the largest flow features~\citep{FBPL2017}. If instead $1<\beta<3$, i.e. for energetic submesoscales, a power-law behavior $\langle R^2(t) \rangle \sim t^{4/(3-\beta)}$ is expected. This is often called a local dispersion regime, because the growth of 
$\langle R^2(t) \rangle$ is in this case driven by velocity differences over lengthscales comparable with the distance between the two particles in a pair~\citep[see, e.g.,][]{lacasce2008}. 
Clearly, this situation includes the well-known Richardson dispersion regime, $\langle R^2(t) \rangle \sim t^3$, corresponding to $E(k) \sim k^{-5/3}$. At even larger times, when the pair-separation distance overcomes the largest eddy size, particles experience uncorrelated velocities and thus relative dispersion follows a slower, standard-diffusion behavior, $\langle R^2(t) \rangle \sim t$.

\begin{figure}
    \centering
    \includegraphics[width=0.9\linewidth]{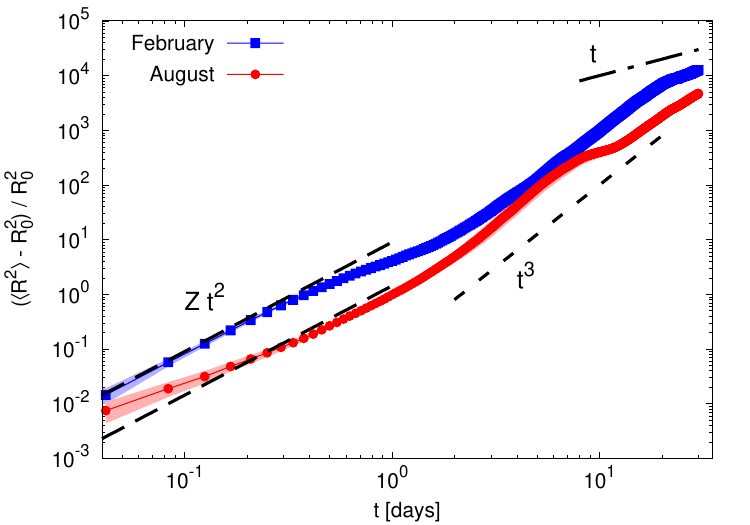}
    \caption{Normalized relative dispersion $(\langle R^2\rangle-R_0^2)/R_0^2$ as a function of time, for February and August in the Kuroshio Extension region. 
    Uncertainties, estimated as the 95\% confidence interval from a bootstrapping procedure, are represented by the shading. 
    }
    \label{fig:f4}
\end{figure}

For the Kuroshio Extension region, relative dispersion as a function of time is shown in Fig.~\ref{fig:f4}, after subtracting the initial value $R_0$ and normalizing by it. At short times, we observe a behavior close to the expected ballistic regime, $\left( \langle R^2 \rangle - R_0^2 \right) / R_0^2 \approx Z t^2$, with $Z$ independently computed from the Eulerian velocity field. The agreement with the theoretical prediction is better in February than in August (for which a slower initial growth is observed) but the prediction gives the right magnitude for both seasons.
We do not have an interpretation of this deviation from the ballistic behavior but remark that it only concerns a time range when the uncertainty on relative dispersion is also larger. 
The larger values of relative dispersion at short times, and hence of enstrophy, in winter than in summer align with the observation of generally more energetic small-scale flows in this season (see Fig.~\ref{fig:f2}). At intermediate times ($1~\text{days}<t<10~\text{days}$), in February, $\langle R^2(t)\rangle$ follows a behavior not far from the Richardson $t^3$ law, before a transition to a linear, diffusive scaling at larger times.
In August, within the same intermediate time range, relative dispersion increases more rapidly (with a slightly steeper slope) before eventually transitioning to what appears to be a $t^3$ scaling. 
If in terms of dispersion regimes the resulting picture qualitatively agrees with the spectra shown in Fig.~\ref{fig:f3}, from a quantitative point of view the situation is less clear. 
%%%%%%%%%%%%%%%%%
% FIG. 4
%%%%%%%%%%%%%%%%%

\begin{figure}
    \centering
    \includegraphics[width=0.9\linewidth]{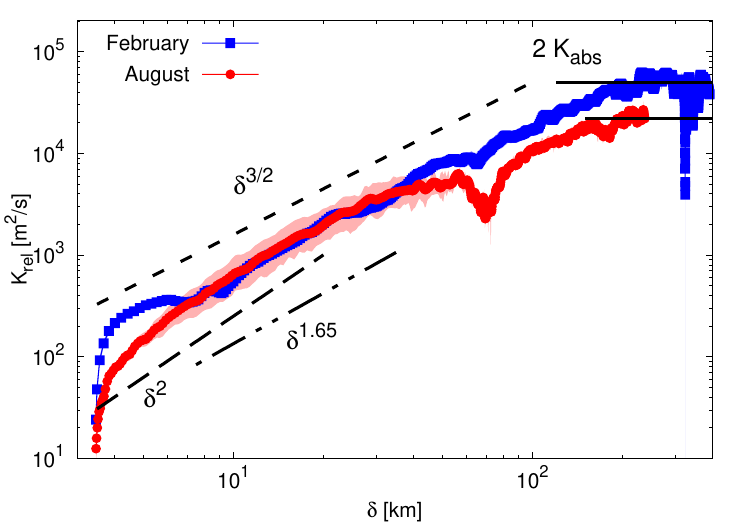}
    \caption{
    Relative diffusivity $K_{rel}$ as a function of the separation distance $\delta=\langle R^2(t)\rangle^{1/2}$, for February and August in the Kuroshio Extension region.  
    The $\delta^{3/2}$ (short-dashed line), $\delta^{1.65}$ (dashed-dotted line) and $\delta^2$ (long-dashed line) behaviors correspond to $\beta = 2$, $\beta=2.3$ and $\beta > 3$, respectively. The horizontal black solid lines represent twice absolute diffusivity at large times, $2 K_{abs}$ (in each month). Uncertainties are estimated as the 95\% confidence interval from a bootstrapping procedure. 
    }
    \label{fig:f5}
\end{figure}

A connected metric of dispersion is relative diffusivity,
%%%%%%%%%%%%%%%%%%%
\begin{equation}
K_{rel} = \frac{1}{2} \frac{d \langle R^2(t) \rangle}{dt}.
\label{eq:reldiff}
\end{equation}
%%%%%%%%%%%%%%%%%%%
While clearly by definition $K_{rel}$ is still a function of time, it is often useful to plot it as a function of the distance $\delta=\langle R^2(t) \rangle^{1/2}$. The results are shown in Fig.~\ref{fig:f5}. 
We preliminarily remark that at the largest separations  [$\delta>O(100)$~km], relative diffusivity approaches a constant value, as expected. In this range, one finds that indeed $K_{rel} \approx 2 K_{abs}$, where $K_{abs}$ is absolute diffusivity (not shown). 
In February, at intermediate scales ($10~\text{km} < \delta < 100~\text{km}$), $K_{rel}$ quite closely follows a $\delta^{3/2}$ scaling, 
indicative of a local dispersion regime, and 
corresponding to a kinetic energy spectrum $E(k)\sim k^{-2}$, in agreement with the measured one (Fig.~\ref{fig:f3}). 
In August, for scales between approximately $10$ and $40$~km, relative diffusivity behaves similarly to the February scaling. In this range, taking into account uncertainties, it is not possible to distinguish between this behavior and the $\delta^{1.65}$ behavior corresponding to the spectral slope $\beta=2.3$. Then $K_{rel}$ decreases for increasing $\delta$, in agreement with Fig.~\ref{fig:f4} where a slow-down of relative dispersion can be seen at around 10~days. This seems to be associated with a change in the regime of growth of $\langle R^2(t)\rangle$, which might be due to efficient particle retention in mesoscale eddies. More importantly, when approaching submesoscales (particularly for $ \delta < 20~\text{km}$), we observe a  tendency towards a steeper growth, compatible with $K_{rel} \sim \delta^2$.
The latter behavior points to nonlocal dispersion and, dimensionally, it corresponds to a smooth flow with $\beta>3$. Therefore, it is at odds with the spectral slope $\beta=2.3$ measured in summer (Fig.~\ref{fig:f3}), a fact that deserves further investigation by means of other indicators.  
%%%%%%%%%%%%%%%%%
% FIG. 5
%%%%%%%%%%%%%%%%%

\begin{figure}
    \centering
    \includegraphics[width=0.9\linewidth]{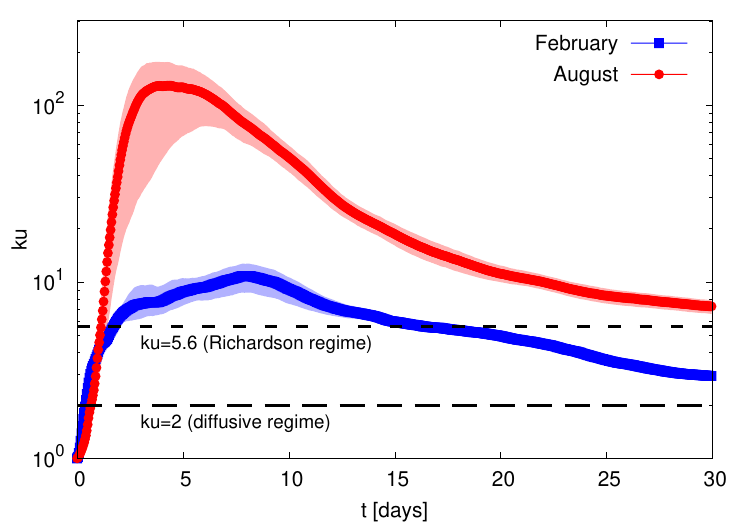}
    \caption{
    Kurtosis of separation $ku$ as a function of time, for February and August in the Kuroshio Extension region. 
    Uncertainties are estimated as the 95\% confidence interval from a bootstrapping procedure. 
    }
    \label{fig:f6}
\end{figure}

Another diagnostic, equally based on a straightforward fixed-time analysis, useful to discriminate between different dispersion regimes, is the kurtosis of the probability density function (pdf) of the pair separation distance~\citep{lacasce2008,LaCasce2010,FBPL2017},
%%%%%%%%%%%%%%%%%%%
\begin{equation}
ku(t) = \frac{\langle R^4(t) \rangle}{\langle  R^2(t) \rangle^2}.
\label{eq:kurtosis}
\end{equation}
%%%%%%%%%%%%%%%%%%%
In a nonlocal dispersion regime, the kurtosis is expected to display fast, exponential growth. For local dispersion, it should level off around a constant value over a finite interval of time [e.g., $ku(t) = 5.6$ for Richardson dispersion]. At larger times, in the diffusive regime, one expects $ku(t)=2$. As compared to relative dispersion and diffusivity, in the kurtosis temporal evolution the differences between winter and summer are much more evident  (Fig.~\ref{fig:f6}).
At short times, the kurtosis grows to values an order of magnitude larger in August than in February, following a quasi-exponential regime. In February, after a rapid increase, kurtosis attains an almost constant plateau at around 15~days, with a value close to  $ku = 5.6$ , the Richardson expectation, before decreasing. These observations then support those from relative diffusivity, suggesting that in winter dispersion is local, while in summer it is nonlocal. 
%%%%%%%%%%%%%%%%%
% FIG. 6
%%%%%%%%%%%%%%%%%

The computation of the previous diagnostics requires performing averages over pairs at any given time along particle trajectories. It is known that such a procedure has some drawbacks, due to the fact that dispersion regimes change in correspondence with lengthscales, not temporal ones~\citep{berti2011lagrangian,CV2013}. As a consequence, fixed-time statistics may be biased by the superposition of different behaviors, due to distinct pairs experiencing different dispersion regimes at the same, common time. 
Fixed-scale analyses, based on computing statistics as a function of the length scales, instead, allow disentangling different dispersion regimes~\citep[see][for a review]{CV2013}. Therefore, we will now  consider the finite-size Lyapunov exponent (FSLE)~\citep{aurell1997predictability,ABCCV1997}, namely a scale-by-scale dispersion rate defined as 
%%%%%%%%%%%%%%%%%%%
\begin{equation}
    \lambda(\delta) =  \frac{\ln{r}}{\langle \tau(\delta) \rangle},
    \label{eq:fsle}
\end{equation}
%%%%%%%%%%%%%%%%%%%
where the average is over all particle pairs and $\tau(\delta)$ is the time needed for the separation distance to grow from $\delta$ to a scale $r \delta$ (with $r > 1$).
As for relative dispersion, dimensional arguments allow to link the FSLE behavior and the kinetic energy spectrum of the underlying flow. In a nonlocal dispersion regime, corresponding to a spectral exponent $\beta>3$ and exponential particle separation, the FSLE should be independent of $\delta$. Its constant value provides an estimate of the maximum Lagrangian Lyapunov exponent and should be proportional to $Z^{1/2}$. For more energetic small-scale flows, when $1<\beta<3$, dispersion is local and the FSLE scales as $\lambda(\delta) \sim \delta^{(\beta-3)/2}$. In particular, Richardson dispersion ($\beta=5/3$) translates into $\lambda(\delta) \sim \delta^{-2/3}$. Finally, in the diffusive regime, holding for separations larger than the largest eddies, one expects $\lambda(\delta) \sim \delta^{-2}$.  

%%%%%%%%%%%%%%%%%
% FIG. 7
\begin{figure}[b!]
    \centering
    \includegraphics[width=0.9\linewidth]{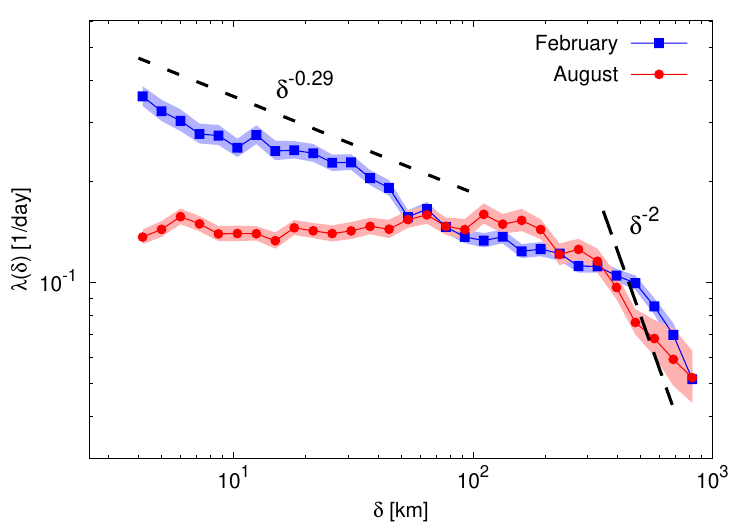}
    \caption{
    FSLE $\lambda(\delta)$ 
    for February and August in the Kuroshio Extension region. The $\delta^{-0.29}$ scaling behavior (short-dashed line), from a fit in the range $4$~km$\leq \delta \leq 100$~km, corresponds to the spectral slope $\beta\simeq 2.4$ 
    and the $\delta^{-2}$ scaling law (long-dashed line) to the diffusive limit. 
    Uncertainties are estimated as the 95\% confidence interval from a bootstrapping procedure. 
    }
    \label{fig:f7}
\end{figure} 
%%%%%%%%%%%%%%%%%

In February, from the smallest sampled separations up to $\delta \simeq 100$~km, the FSLE follows the scaling $\delta^{-\gamma}$, with $\gamma \simeq 0.29$ from a fit between $\delta=4$ and $100$~km (Fig.~\ref{fig:f7}), further supporting the indication of local dispersion, associated with energetic submesoscales. 
From the value of the exponent $\gamma$ one has $\beta \simeq 2.4$, larger than the mean value ($\beta=2$) of the slope measured from the spectrum, but compatible with its upper bound. 
In contrast, in August, in the same range of scales ($5~\text{km} \lesssim \delta \lesssim 100~\text{km}$), $\lambda(\delta)$ is virtually independent of $\delta$. 
This confirms, once more, the essentially nonlocal character of dispersion in this season, in spite of the spectrum [$E(k) \sim k^{-2.3}$] being shallower than $k^{-3}$. 
Finally, in both winter and summer, the FSLE eventually approaches a diffusive regime, indicated by a $\delta^{-2}$ behavior, for $\delta \gtrsim 300$~km. The latter scale is in reasonable agreement with the size of the largest eddies, $\ell_M \sim 1/k_M \approx 200$~km, estimated from the wavenumber $k_M$ where the kinetic energy spectra peak (Fig.~\ref{fig:f3}). 

Summarizing, the picture emerging from this analysis indicates that seasonality has an important role on Lagrangian dispersion in this region. In particular, the overall coherence, in each season, of the different metrics considered highlights that in winter (February) dispersion is local, while in summer (August) it is nonlocal. In winter, the scaling behaviors of the Lagrangian diagnostics tend to align with the usual  predictions from turbulence theory based on the slope of the kinetic energy spectrum. Specifically, to reasonable extent, they match the dimensional expectations based on a power-law decay of a kinetic energy spectrum with an exponent $\beta \gtrsim 2$, as the one measured from the Eulerian velocity field. Relative dispersion is the only exception, presenting a slightly different scaling perhaps more compatible with $\beta=5/3$, which is however not too far from the value estimated from other indicators. In summer, the kinetic energy spectrum has a (clearer) slope $\beta \simeq 2.3$, which would predict local dispersion, in contrast with the Lagrangian results.

We conclude this section by noting that these findings appear 
in line with the visual inspection of Fig.~\ref{fig:f1}, illustrating how particles disperse in the flow. After one month of simulation, Lagrangian particles tend to accumulate along fronts and inside large-scale vortices in summer (Fig.~\ref{fig:f1}d) while they are more efficiently homogenized through the domain and at all scales in winter (Fig.~\ref{fig:f1}b). Such a difference hints at Lagrangian transport driven by mesoscale fronts and eddies (i.e. nonlocal dispersion) in summer, and at smaller-scale fronts and eddies tending to disperse particles through the flow (as under local dispersion) in winter. 

%%%%%%%%%%%%%%%%%%%%%%%%%%%%%%%%%%%%%%%%%%%%%%
\section{Lagrangian dispersion interpretation based on a slow-fast flow decomposition}\label{sec:llc_igw_bm}

The results in the previous section indicate that, in summer, there is a clear disagreement between relative dispersion indicators and their predictions from the kinetic energy spectrum, contrary to what one would expect within the theory of QG turbulence. Therefore, one question arises: what is the origin of such disagreement? 
%%%%%%%%%%%%%%%%%
% FIG. 8
\begin{figure}[h!]
    \centering
    \includegraphics[width=0.9\linewidth]{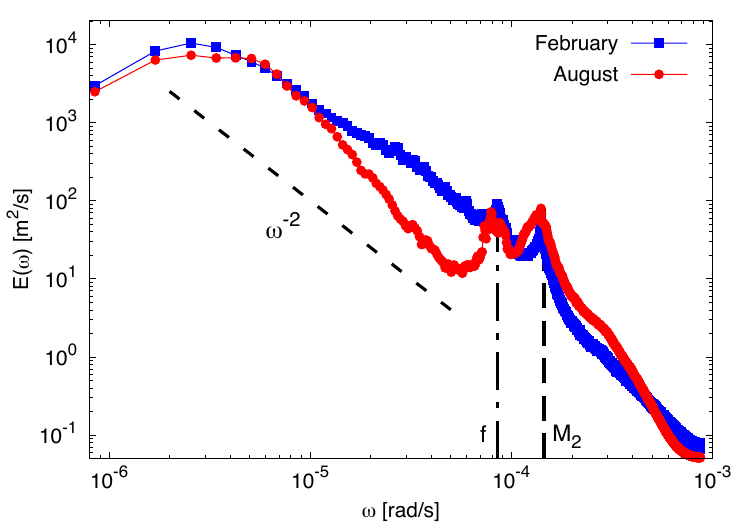}
    \caption{Lagrangian frequency spectra of kinetic energy $E(\omega)$ for February and August in the Kuroshio Extension region. 
    The vertical lines indicate the Coriolis ($f$) and semidiurnal tidal ($M_2$) frequencies.
    }
    \label{fig:f8}
\end{figure}
%%%%%%%%%%%%%%%%%

%%%%%%%%%%%%%%%%%%%%%%%%%%%%%%%%%%%%%%%%
\subsection{Lagrangian frequency spectra}\label{sec:llc_lagr_spec}

One candidate to answer the above question is the presence of IGWs. A first way to determine their importance for the Lagrangian dynamics is to compute the Lagrangian frequency spectrum of kinetic energy $E(\omega)$. As observed in Fig.~\ref{fig:f8}, for both February and August, the spectra peak at low frequencies, suggesting that the advection of Lagrangian particles is governed by slow 
(presumably quasi-balanced) motions. One can also clearly distinguish two peaks, corresponding to the Coriolis ($f$) and tidal ($M_2$) frequencies with periods $T_f\approx 20.53$~h and $T_{M_2}\approx12.65$~h, respectively.
In August, these peaks (most likely  associated with IGWs) are more pronounced and constitute a significant part of the Lagrangian energy. This result highlights the fact that Lagrangian trajectories are  sensitive to the high-frequency components of the flow. In February,  on average, the scaling of the spectrum is not far from $\omega^{-2}$, which corresponds to an exponential decay of the velocity autocorrelation function (not shown). 

%%%%%%%%%%%%%%%%%%%%%%%%%%%%%%%%%%%%%%%%

\subsection{Frequency-wavenumber energy spectra}\label{sec:llc_wkspec_kuro}

\begin{figure}
    \centering
    \includegraphics[width=0.45\textwidth]{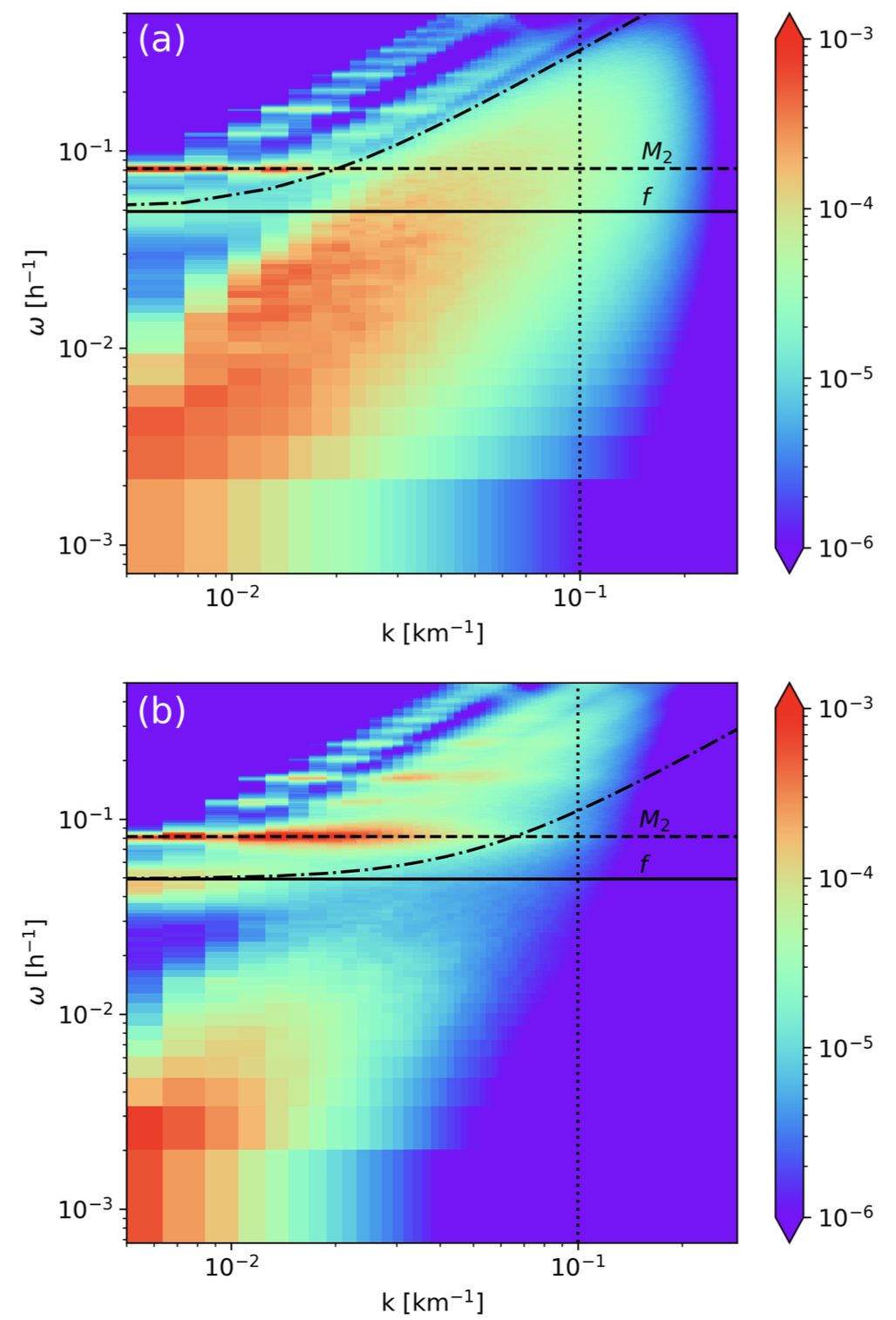}
    \caption{Frequency-wavenumber spectra of kinetic energy $E(k,\omega)$ in the Kuroshio Extension region during February (a) and August (b); here  spectra are shown in variance-preserving form $k \, \omega \, E(k,\omega)$ with units in m$^2$~s$^{-2}$. 
    The horizontal solid and dashed lines indicate the Coriolis ($f$) and semidiurnal tidal ($M_2$) frequencies, respectively, while the dashed-dotted line shows the dispersion-relation curve for the $10^\mathrm{th}$ baroclinic mode. The corresponding deformation radii are $L_R=65$~km (a) and $L_R=20$~km (b). 
    }
    \label{fig:f9}
\end{figure}

We next analyze the respective contributions of M/SM motions and IGWs to the Eulerian kinetic energy spectrum. Following the methodology of \citet{Torresetal_2018,Torresetal_2022}, we compute the frequency-wavenumber ($\omega-k$) spectrum of kinetic energy, which is shown in Fig.~\ref{fig:f9}a for February and in Fig.~\ref{fig:f9}b for August. The distinction between M/SM and IGWs can be made using the dispersion-relation curve of IGWs,  $\omega^2 = c^2 k^2 + f^2$~\citep{Torresetal_2018}. Here $c$, $k$, and $f$ are, respectively, the phase speed of inertio-gravity waves, the isotropic horizontal wavenumber, and the Coriolis frequency. This relation can be reformulated to incorporate the deformation radius $L_R\approx c/|f|$, leading to $\omega^2 = f^2\,(L_R^2 k^2+1)$~\citep{Sutherland_2010}. As seen in Fig.~\ref{fig:f9}, using the dispersion relation for the $10$\textsuperscript{th} vertical mode
(dashed-dotted line) allows to make a clear distinction between IGWs and balanced, M/SM motions. Indeed, this mode corresponds to the highest baroclinic mode resolved in the LLC4320 simulation and, hence, is the most relevant one for partitioning the flow into balanced and higher-frequency, wavy motions~\citep{Torresetal_2018}. In this region, the value of $L_R$ is $\simeq 65$~km in winter and $\simeq 20$~km in summer.
This partitioning method is essential because IGWs and high-frequency submesoscales share similar frequencies, making it difficult to distinguish between them using simpler techniques, as e.g. filtering based solely on frequencies, such as $f$ or $M_2$~\citep{jones2023using}. 

In winter (Fig.~\ref{fig:f9}a), the energy is concentrated at frequencies below those of IGWs, while internal tides and inertial motions do not seem to contribute to it significantly. This suggests that the energy is essentially all contained in M/SM motions. 
In summer (Fig.~\ref{fig:f9}b), the energetic content of high-frequency IGWs increases, 
with a marked concentration of energy around $M_2$, while that of submesoscales considerably decreases. This is in line with the Lagrangian energy spectrum (Fig.~\ref{fig:f8}), for which we observe a spectral gap between the energetic low-frequencies and the inertial and semidiurnal motions. 
%%%%%%%%%%%%%%%%%
% FIG. 9
%%%%%%%%%%%%%%%%%

\begin{figure}
    \centering
    \includegraphics[width=0.9\linewidth]{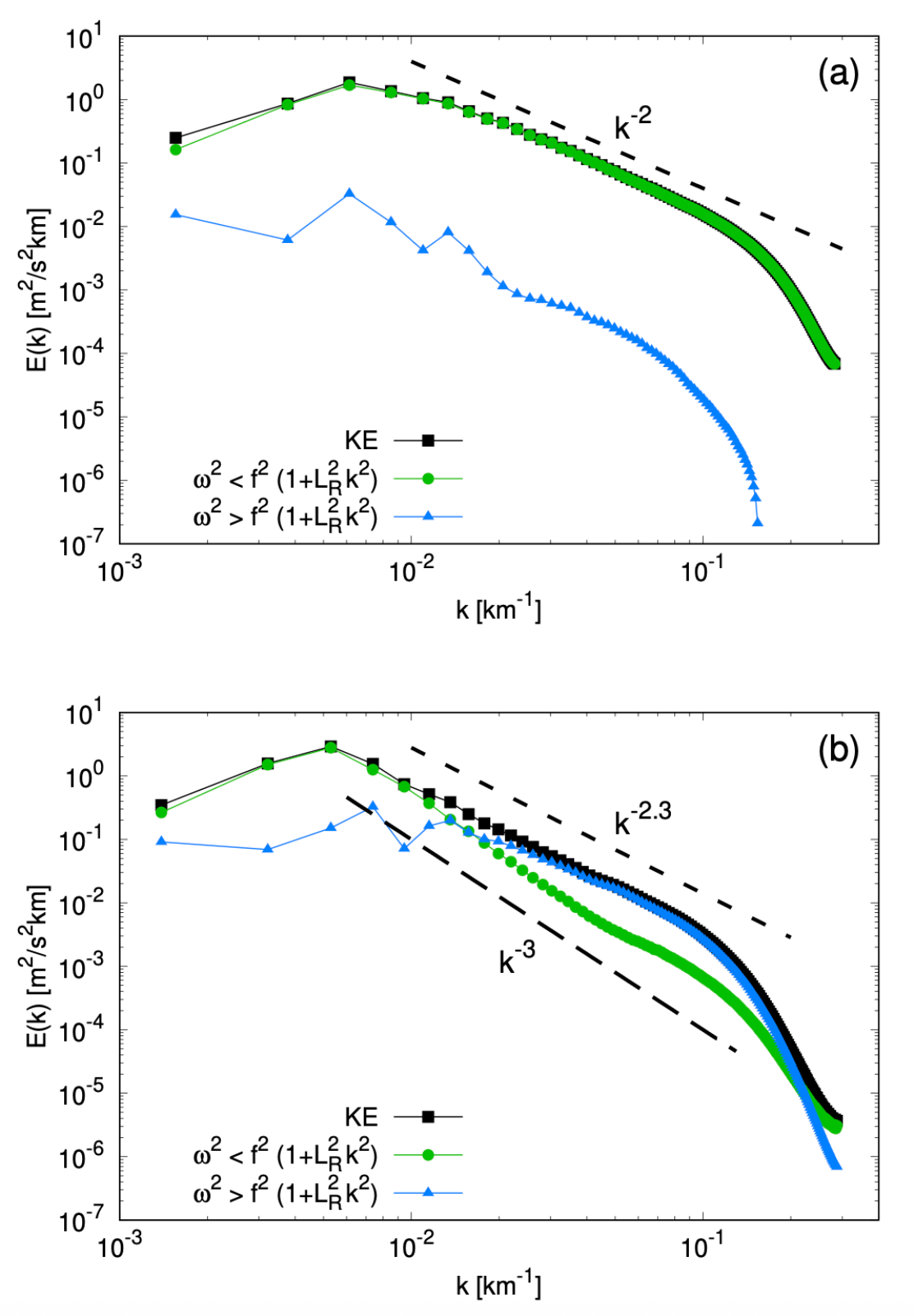}
    \caption{Decomposition of the kinetic energy wavenumber spectra $E(k)$ for February (a) and August (b) in the Kuroshio Extension region. 
    The spectrum of the total kinetic energy (KE) is shown by black square points. The contribution from frequencies such that $\omega^2<f^2(1+L_R^2 k^2)$ corresponds to the green dots, while the blue triangles are for frequencies $\omega^2>f^2(1+L_R^2 k^2)$. The corresponding deformation radii are $L_R=65$~km (a) and $L_R=20$~km (b). The reference lines $k^{-2}$ in (a), $k^{-3}$ and $k^{-2.3}$ in (b), are also shown for comparison.
    }
    \label{fig:f10}
\end{figure}

From the frequency-wavenumber spectrum $E(k,\omega)$, we can evaluate the IGW contributions to the wavenumber spectrum of kinetic energy $E(k)$ by integrating $E(k,\omega)$ over frequencies satisfying only either $\omega^2 < f^2 (1 + L_R^2 k^2)$ or $\omega^2 > f^2 (1 + L_R^2 k^2)$. This procedure reveals that in February (Fig.~\ref{fig:f10}a) IGWs are less energetic than M/SM motions by two orders of magnitude. The latter, then, indeed account for most of the kinetic energy in the surface flow at all scales: the associated spectrum is almost identical to that of the total kinetic energy, and both appoximately follow a $ k^{-2} $ scaling.  In August (Fig.~\ref{fig:f10}b), instead, we observe that at small wavenumbers (lengthscales $ >100 $~km), mesoscale motions still dominate, but at larger wavenumbers (lengthscales $ <50 $~km), submesoscales become less energetic and IGWs provide the leading contribution to the kinetic energy spectrum. The small-scale IGW spectrum scales as $k^{-2.3}$ , 
while that of low-frequency (M/SM) motions behaves as $k^{-3}$ up to $k=0.04$~km$^{-1}$. Such steeper spectrum (from M/SM) corresponds theoretically to a regime of nonlocal particle dispersion. This result is thus consistent with IGWs having little to no effect on relative dispersion, despite having a prominent signature on the small-scale energetic content of the flow. 
%%%%%%%%%%%%%%%%%
% FIG. 10
%%%%%%%%%%%%%%%%%

%%%%%%%%%%%%%%%%%%%%%%%%%%%%%%%%%%%%%%%%%%%%%%%%%%%%%%%%%%%
\subsection{Decomposition into rotational and divergent motions}\label{sec:llc_slow-fast_kuro}

To obtain a finer picture of what dynamical processes affect Eulerian spectra,  kinetic energy can be decomposed into rotational (KE$_\zeta$) and divergent (KE$_\Delta$) components, using Helmholtz decomposition~\citep{Buhler_Callies_Ferrari_2014,Rocha2016,Torresetal_2018}: 
%%%%%%%%%%%%%%%%%%%
\begin{equation}
    KE_\zeta(k) = \frac{1}{2}\int \frac{\left|\hat{\zeta}(k,\omega)\right|^2}{k^2}\,d\omega 
    \label{eq:llc:ke_zeta}
\end{equation}
%%%%%%%%%%%%%%%%%%%
and
%%%%%%%%%%%%%%%%%%%
\begin{equation}
    KE_\Delta(k) = \frac{1}{2}\int \frac{\left|\hat{\Delta}(k,\omega)\right|^2}{k^2}\,d\omega\,,
    \label{eq:llc:ke_Delta}
\end{equation}
%%%%%%%%%%%%%%%%%%%
where $\hat{\zeta}(k,\omega)$ and $\hat{\Delta}(k,\omega)$ are the spatiotemporal Fourier transforms of vorticity $\zeta$ and divergence $\Delta=\partial_x u + \partial_y v$, respectively.
Mesoscale motions are typically close to geostrophic balance and hence nondivergent. On the other hand, in general, both submesoscales (induced by frontal dynamics) and IGWs 
contribute to the divergence field. 
We then further separate each component into $KE_{\zeta,\Delta}^-$, representing low-frequency processes such that $\omega^2 < f^2 (1 + L_R^2 k^2) $, and $KE_{\zeta,\Delta}^+$, representing high-frequency processes such that $ \omega^2 > f^2 (1 + L_R^2 k^2) $. 

\begin{figure*}
    \centering
    \includegraphics[width=0.9\linewidth]{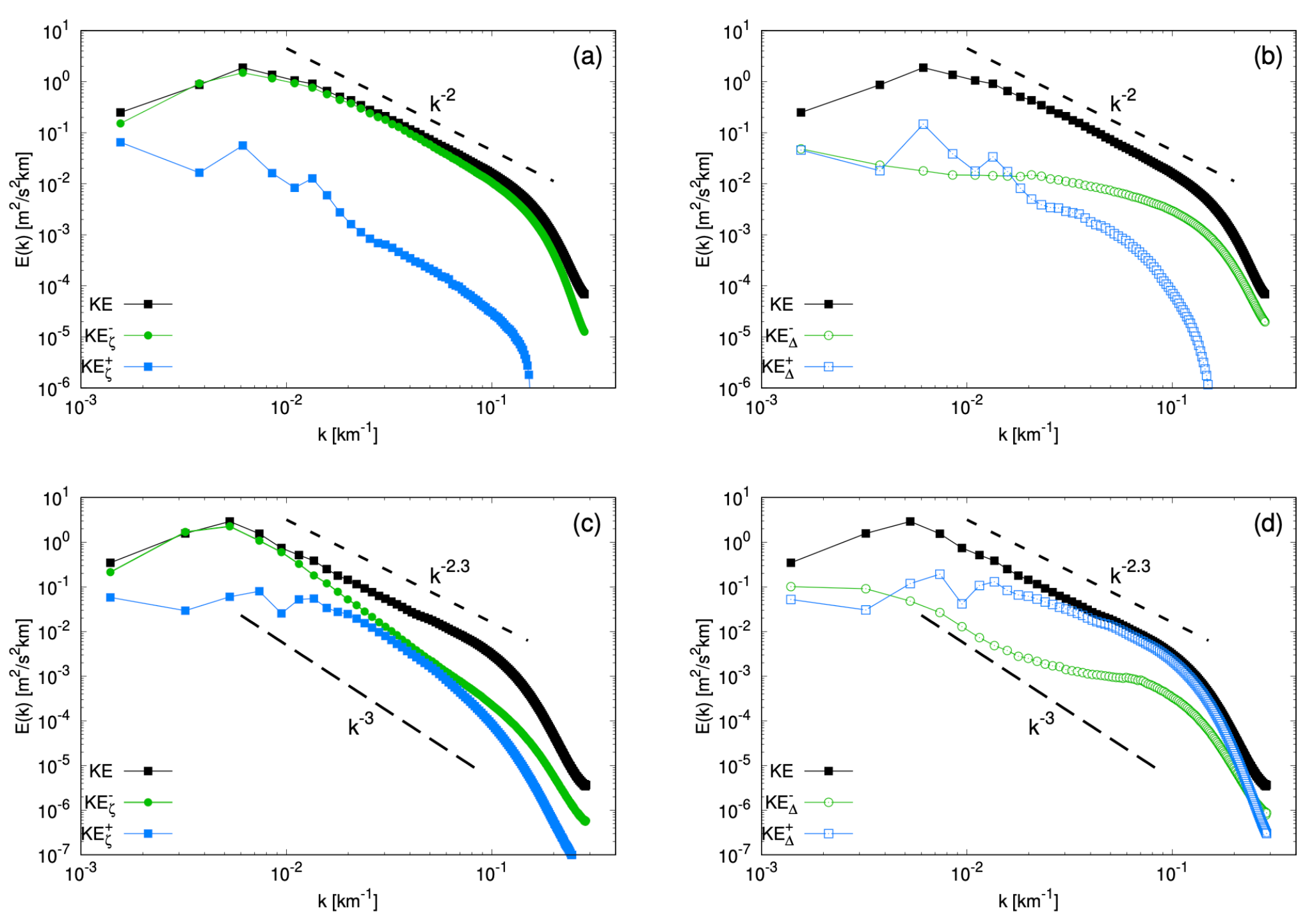}
    \caption{Wavenumber spectra of kinetic energy $ E(k)$ in the Kuroshio Extension region. The spectrum of the total kinetic energy (KE) is shown by black square points. (a, c) Spectra of the rotational component KE$_\zeta$; (b, d) spectra of the divergent component $KE_{\Delta}$. In each case, the flow is further partitioned into low and high-frequency motions as in Fig.~\ref{fig:f10}. 
    Panels (a) and (b) correspond to February, panels (c) and (d) to August. 
    }
    \label{fig:f11}
\end{figure*}

Figure~\ref{fig:f11} shows the results of this partitioning for February [panels (a) and (b)] and August [panels (c) and (d)]. In February, the flow is dominated by its rotational component, primarily from M/SM motions (Fig.~\ref{fig:f11}a). At all scales, the divergent component from both M/SM and IGWs contributes little to the overall kinetic energy (Fig.~\ref{fig:f11}b). 
In August, the situation is different. At low wavenumbers [lengthscales $ > (50-100) $~km], rotational M/SM motions dominate (Fig.~\ref{fig:f11}c), while at higher wavenumbers the divergent contribution from IGWs becomes dominant in the kinetic energy spectrum (Fig.~\ref{fig:f11}d). 
Notably, the spectrum of slow motions associated with vorticity $KE_{\zeta}^-$ has, in this season, a clear $ k^{-3} $ scaling over an extended wavenumber range. The corresponding spectrum of fast IGWs $KE_{\zeta}^+$ is generally shallower, with values smaller than those of $KE_{\zeta}^{-}$, except in a narrower range of scales where it is comparable (and behaves similarly) to $KE_{\zeta}^-$. These results clearly show that the full wavenumber kinetic energy is not necessarily representative of the balanced dynamics. 
%%%%%%%%%%%%%%%%%
% FIG. 11
%%%%%%%%%%%%%%%%%

%%%%%%%%%%%%%%%%%%%%%%%%%%%%%%%%%%%%%%%%%%%%%%
\section{Comparison with results in the Gulf Stream region}\label{sec:llc_gs}

In order to test the generality of the results in Sec.~\ref{sec:llc_lagrangian} and Sec.~\ref{sec:llc_igw_bm}, here we provide a discussion of the main picture emerging from the same approach in another energetic region, close to the Gulf Stream. Its exact location and a more extensive characterization of the Eulerian and Lagrangian properties for this case study are reported in the Appendix. 

\begin{figure}
    \centering
    \includegraphics[width=0.9\linewidth]{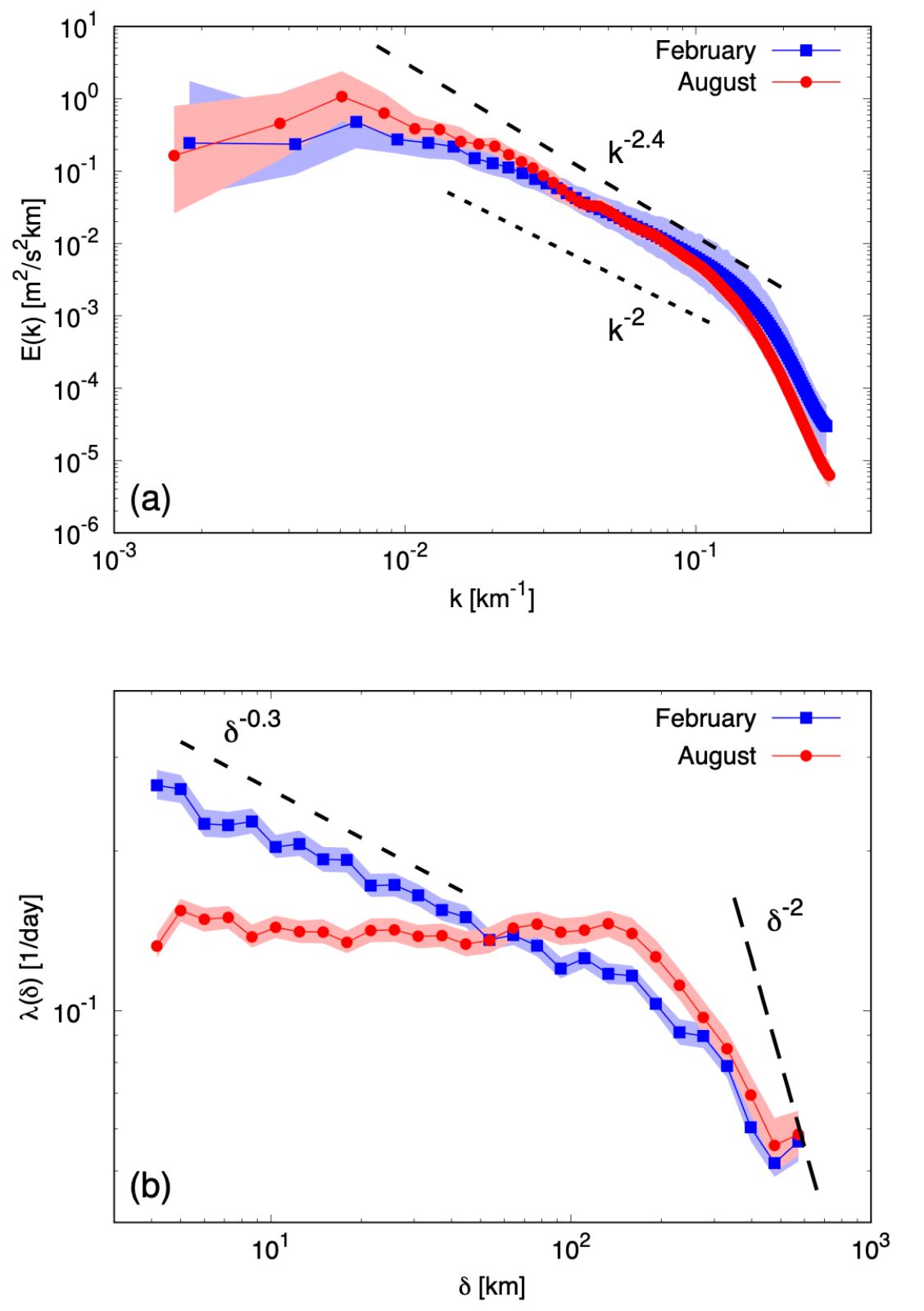}
    \caption{(a) Wavenumber spectra of horizontal kinetic energy in the Gulf Stream region, averaged over February (blue squares) and August (red dots). For both months, the shaded areas represent the temporal variability of the spectrum.
    (b) Corresponding FSLE $\lambda(\delta)$ in the same region and for the same months. The $\delta^{-0.3}$ scaling law (short-dashed line) corresponds the spectrum $E(k) \sim k^{-2.4}$, while the $\delta^{-2}$ scaling law (long-dashed line) corresponds the diffusive limit. 
    Uncertainties are estimated as the 95\% confidence interval from a bootstrapping procedure. 
    } 
    \label{fig:f12}
\end{figure}

As in Kuroshio Extension, the wavenumber kinetic energy spectrum (Fig.~\ref{fig:f12}a), is in both seasons quite energetic at submesoscales. We note, however, that in this region the February and August spectra are remarkably close (indeed, they are equal, within error bars) and scale approximately as $E(k) \sim k^{-2.4}$ over more than a decade. The summer spectrum is a bit more energetic and steeper at large scales, while the winter one is slightly shallower, with a slope also compatible with $\beta=2$ over a shorter wavenumber subrange. To quantify the scale-by-scale intensity of the pair-dispersion process, we focus on the FSLE (Fig.~\ref{fig:f12}b). The power-law and constant behaviors in February and August, respectively, clearly indicate that dispersion is local in winter and nonlocal in summer. Interestingly, from a quantitative point of view, we observe here the same season-dependent agreement with the predictions from energy spectra as in Kuroshio Extension. Indeed, the winter scaling $\lambda(\delta) \sim \delta^{-0.3}$ quite nicely matches the spectrum-based expectation  $\lambda(\delta) \sim \delta^{(\beta-3)/2}$ (with $\beta=2.4$), but the extended plateau, $\lambda(\delta) \simeq \mathrm{const}$, found in summer is in evident contrast with the corresponding spectrum, which would even indicate a different dispersion regime (local rather than nonlocal). 
%%%%%%%%%%%%%%%%%
% FIG. 12
%%%%%%%%%%%%%%%%%

\begin{figure}
    \centering
    \includegraphics[width=0.45\textwidth]{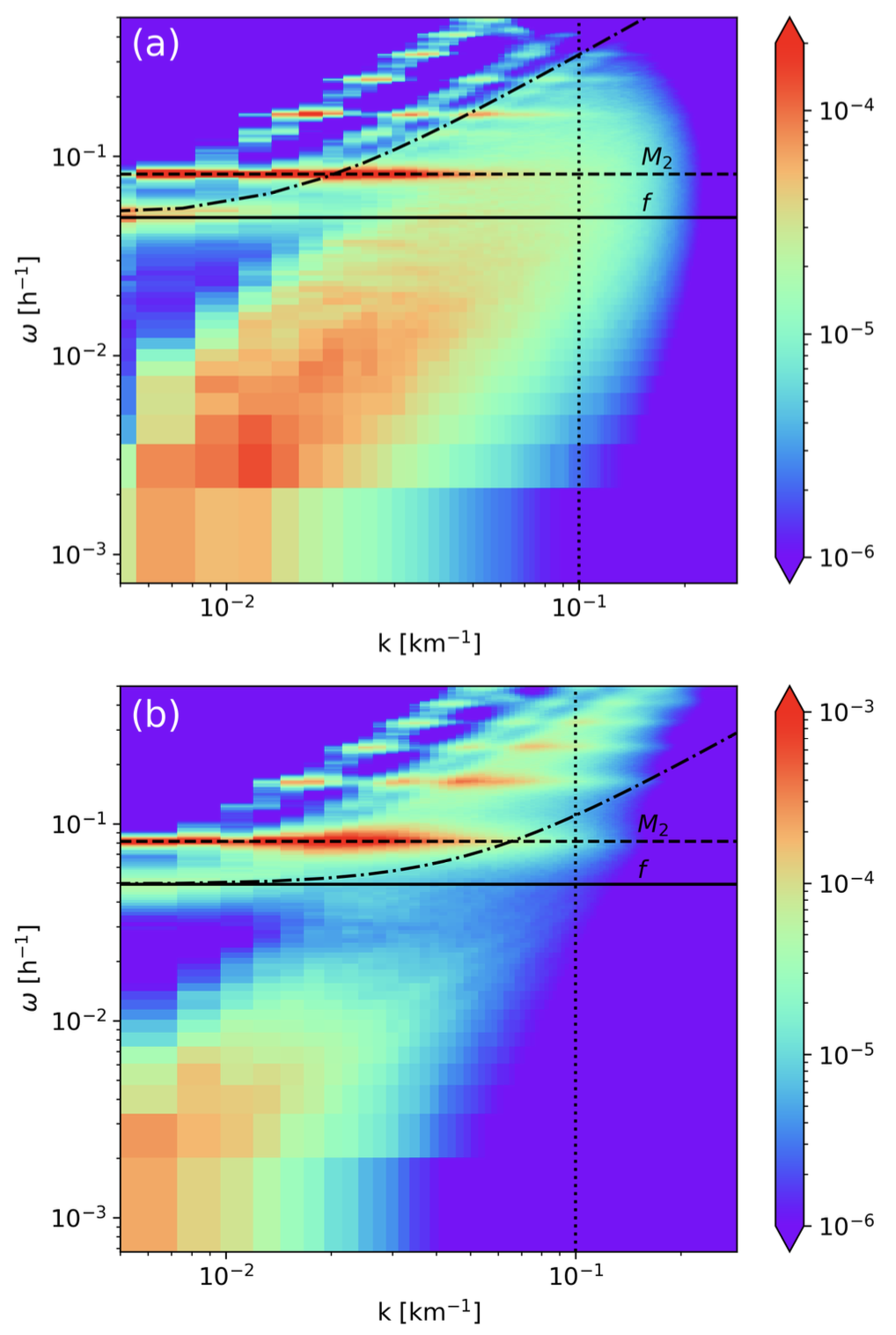}
    \caption{Frequency-wavenumber spectra of kinetic energy $E(k,\omega)$, defined as in Fig.~\ref{fig:f9}, in the Gulf Stream region during February (a) and August (b). For the (dashed-dotted) curve representing the dispersion relation, the deformation radii are $L_R=65$~km (a) and $L_R=20$~km (b). 
    }
    \label{fig:f13}
\end{figure}

As before, we then resort to frequency-wavenumber kinetic energy spectra to assess the relative importance of high and low frequency motions in each season (Fig.~\ref{fig:f13}). 
The global picture returned by such spectra is very similar to the one found in Kuroshio Extension, which also confirms that these two energetic regions share the same qualitative dynamical features. Specifically, M/SM motions dominate the energetic content of the flow in February; in summer IGWs are considerably more energetic than in winter, and in parallel the intensity of the flow at submesoscales gets reduced. Minor quantitative differences among the two regions can  also be noticed. 
For instance, here the flow is less energetic, particularly in the submesoscale range in winter (as also observed from the slightly steeper February wavenumber spectrum), with respect to that found in Kuroshio Extension. 
%%%%%%%%%%%%%%%%%
% FIG. 13
%%%%%%%%%%%%%%%%%

\begin{figure}
    \centering
    \includegraphics[width=0.9\linewidth]{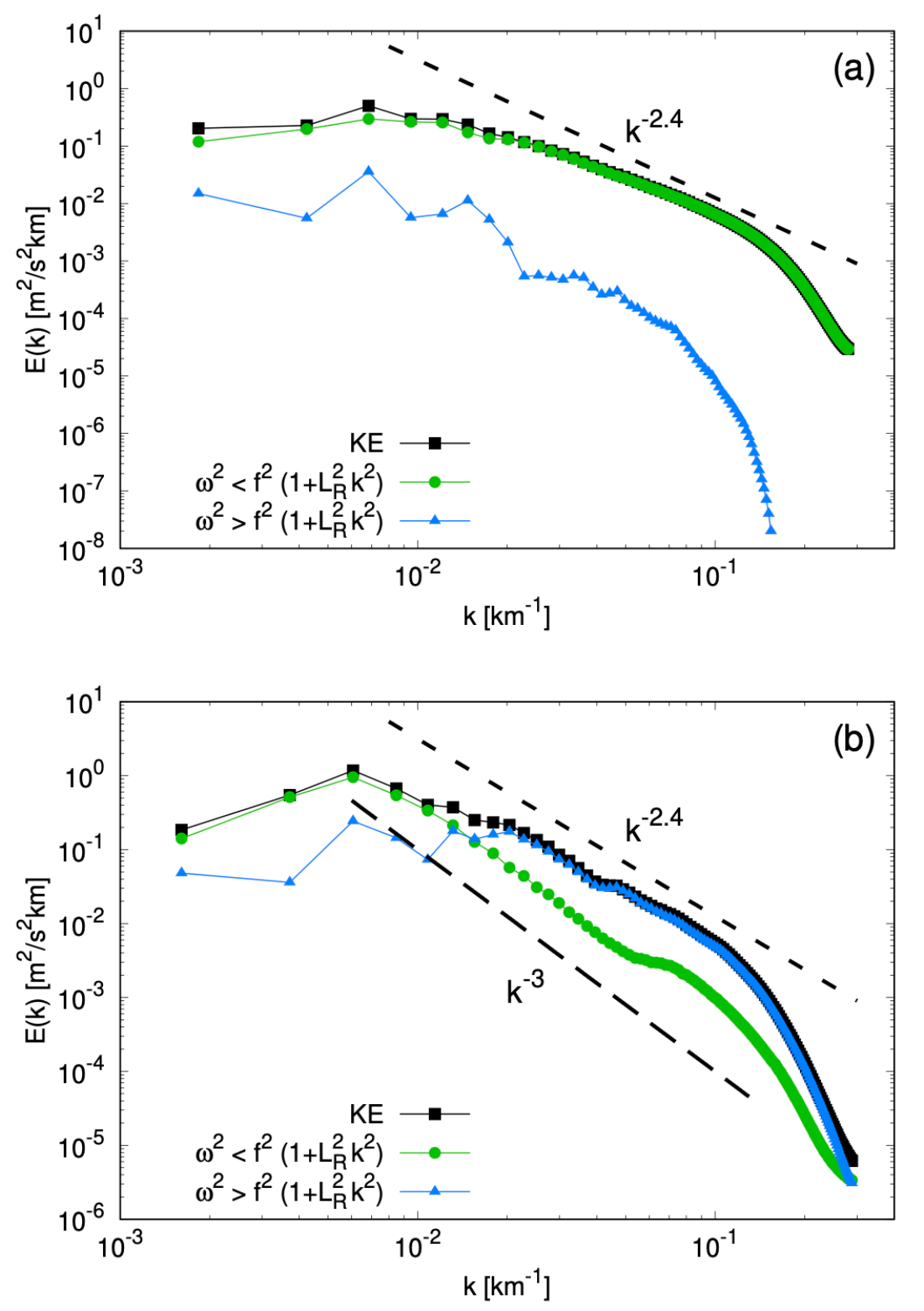}
    \caption{Decomposition of the wavenumber spectra of kinetic energy $E(k)$, as in Fig.~\ref{fig:f10}, for February (a) and August (b) in the Gulf Stream region. 
    The reference lines $k^{-2.4}$ in (a), $k^{-3}$ and $k^{-2.4}$ in (b) are also shown for comparison. The deformation radii are $L_R=65$~km (a) and $L_R=20$~km (b).
    }
    \label{fig:f14}
\end{figure}

Using the spatiotemporal spectra $E(k,~\omega)$, we next compute the contributions of low and high-frequency motions to the total wavenumber kinetic energy spectrum. The results, shown in Fig.~\ref{fig:f14} for both seasons, closely resemble those found in Kuroshio Extension. In winter, M/SM motions [corresponding to frequencies $\omega^2<f^2(1+L_R^2 k^2)$] essentially account for the full kinetic energy at all scales and their spectrum is then close to $k^{-2.4}$. The summer spectrum is dominated by the slow M/SM at scales larger than $50-100$~km and by IGWs [corresponding to frequencies  $\omega^2>f^2(1+L_R^2 k^2)$] at smaller scales. The M/SM spectrum scales as $k^{-3}$, the IGW one as $k^{-2.4}$. 
Therefore, the winter and summer energy spectra, when temporally filtered to retain only the contribution from low-frequency motions, are respectively compatible with the observed local and nonlocal dispersion regimes. This illustrates that the picture found in Kuroshio Extension might be more general and, thus, confirms that while contributing to the kinetic energy spectrum, IGWs are unlikely to have a measurable impact on relative dispersion, at least in the range of separations explored in this study.  
%%%%%%%%%%%%%%%%%
% FIG. 14
%%%%%%%%%%%%%%%%%

%%%%%%%%%%%%%%%%%%%%%%%%%%%%%%%%%%%%%%%%%%%%%%
\section{Conclusions}\label{sec:concl}

We investigated Lagrangian particle transport at the ocean surface, using the high-resolution global-ocean simulation LLC4320, which incorporates internal tides in addition to meso and submesoscale dynamics. We examined in details particle pair dispersion in the Kuroshio Extension in two different seasons. We then extended our analysis to another study region, close to the Gulf Stream. 
The surface velocities from the model were used to advect Lagrangian tracers over the months of February and August (representative of winter and summer, respectively).
The pair-dispersion process was analyzed by means of two-particle statistical indicators, which allow to identify different dispersion regimes and, in principle, to link the Lagrangian results and the Eulerian flow properties via dimensional arguments developed in the framework of QG turbulence~\citep{lacasce2008,FBPL2017}. 

Our findings demonstrate that dispersion is local, meaning controlled by flow features having the same size as the particle separation distance, in winter, and nonlocal, i.e. dominated by the largest flow scales, in summer. This is most clearly revealed by the FSLE, measuring the scale-by-scale dispersion rate, but it is also confirmed by other space or time-dependent diagnostics. In winter, the observed behaviors of Lagrangian indicators, to fair extent, match the dimensional expectations constructed from the slope of the wavenumber kinetic energy spectrum. In summer, however, the predictions based on the spectrum are not confirmed by the actual Lagrangian statistics. The disagreement is not only quantitative but also qualitative: based on the spectrum one would expect local dispersion, while the analysis of particle trajectories indicates that dispersion is nonlocal. 

Examination of spatiotemporal kinetic energy spectra revealed key to understand this apparent discrepancy in summer. 
Computing how energy is distributed among both wavenumbers and frequencies, indeed, allows to separate the contributions from the slower M/SM components of the flow and faster IGWs. Through this approach, we could show that the observed dispersion behaviors reasonably agree with the  predictions based on the wavenumber kinetic energy spectrum associated with the slow, nearly balanced (and mainly rotational) part of the velocity field. In winter high-frequency motions marginally contribute to the spectrum. In summer, they dominate energetically only at scales smaller than roughly $50$~km, and our results are consistent with the dispersion process being controlled by the more intense strain associated with the large-scale, lower-frequency flow, and not the high-frequency one. 
No evidence of an impact of internal waves on pair dispersion was found in the LLC4320 simulation in the Kuroshio Extension region. This picture is further supported by the same analysis conducted in another energetic region, close to the Gulf Stream, sharing similar statistical properties of the Eulerian flow, where we essentially observed the same Lagrangian phenomenology. 

Understanding how general these conclusions are remains an open point, and examining the geographic (and seasonal) variability of relative dispersion appears to us a valuable perspective for future work. To our knowledge, to date only few studies have addressed the impact of IGWs on Lagrangian tracer dispersion, and the conclusions appear varied. For instance, in a study using in-situ and synthetic surface drifters in the Gulf of Mexico~\citep{Beron2016} it was argued that fixed lengthscale indicators, like the FSLE, should be affected by inertial oscillations, which, however, is not the case in our findings. It might then be interesting to correlate the Lagrangian dispersion properties observed in that region with the statistical features of the slow and fast components of the associated Eulerian flow. Another study by~\citet{tranchant2025swot} investigated drifter dispersion in an energetic meander of the Antarctic Circumpolar Current, over a specific period of time, where waves seem to be rather weak. By comparing with virtual drifters advected by SWOT velocities, 
the authors  showed that balanced motions dominate dispersion at scales larger than $\approx 10$~km. Those results, to some extent, align with ours in winter. It would seem to us interesting to complement them with an analysis over different periods and, again, perhaps an examination of spatiotemporal spectra from a high-resolution numerical model. 

Simplified models and PE simulations in smaller domains may also reveal useful to gain further insight into the basic physical mechanisms and to estimate the critical lengthscale below which IGWs may become important for dispersion. Following this approach,~\citet{Wangetal_2018} investigated the destabilization of a circular front in the presence of a wealth on internal waves. While in that case the FSLE is sensitive to inertial oscillations, this effect is only observed at scales smaller than an inertial-oscillation scale $V/f$, where $V$ is the typical velocity of Lagrangian particles. In our case, in August such lengthscale ($4.59$~km) is very close to the first separation value ($\delta \approx 4.17$~km) used in the FSLE computation. 
A similarly crude estimate based on the semidiurnal tidal frequency, $M_2>f$, would give an even smaller typical length. Resolving smaller scales in the Lagrangian dispersion process would require quite a smaller initial pair separation (currently it is $R_0 \approx 3.48$~km). For this, in turn, simulations at even higher spatial resolution than the present ones would be needed, considering that the inertial-oscillation scale is close to LLC4320 horizontal grid spacing, where numerical diffusivity smoothens the flow. These considerations explain, at least qualitatively, why our summer FSLE is insensitive to possible effects due to internal waves. 

We conclude by shortly commenting on the implications of our results for the interpretation of the new, high-resolution altimetry data provided by SWOT. When high-frequency motions are relatively weak, as in our winter situations, the theoretical links between the spectral kinetic energy distribution of the Eulerian flow and relative-dispersion properties should reveal useful to predict the latter. Pending the geostrophic approximation is sufficiently accurate, the satellite-derived velocity field should enable more direct and local predictions of transport and dispersion via Lagrangian advection by the geostrophic velocity field. Note that \citet{yu2021geostrophy} and \citet{demol2025} have quantified the validity of geostrophy at global scales from numerical models and observations, respectively. More studies are required in order to identify general conditions of validity, e.g. in terms of spatial/temporal scales and flow conditions, and hence verify our ability to estimate dispersion properties from SWOT and the nature of the signal processing required to do so. 
However, when internal waves are more important, as in summer in this study, it is unlikely that such theoretical links remain meaningful to obtain information about dispersion, unless high-frequency motions are filtered out from the satellite-derived velocities. Future missions such as Odyssea~\citep{torres2023anticipated} may bring useful complementary information to estimate the low-frequency component of the flow required to assess ocean-surface Lagrangian dispersion.

%\clearpage

%%%%%%%%%%%%%%%%%%%%%%%%%%%%%%%%%%%%%%%%%%%%%%%%%%%%%%%%%%%%%%%%%%%%%
% ACKNOWLEDGMENTS
%%%%%%%%%%%%%%%%%%%%%%%%%%%%%%%%%%%%%%%%%%%%%%%%%%%%%%%%%%%%%%%%%%%%%
\acknowledgments

This work was supported by the French Space Agency CNES (Centre National d’Etudes Spatiales) through the SWOT mission in the framework of DIEGO and DIEGOB projects. The dataset was analyzed on the Datarmor supercomputer at IFREMER in Brest.

%%%%%%%%%%%%%%%%%%%%%%%%%%%%%%%%%%%%%%%%%%%%%%%%%%%%%%%%%%%%%%%%%%%%%
% DATA AVAILABILITY STATEMENT
%%%%%%%%%%%%%%%%%%%%%%%%%%%%%%%%%%%%%%%%%%%%%%%%%%%%%%%%%%%%%%%%%%%%%
% 

\datastatement

The data that support the findings of this study are available
from the corresponding author upon reasonable request. 

%%%%%%%%%%%%%%%%%%%%%%%%%%%%%%%%%%%%%%%%%%%%%%%%%%%%%%%%%%%%%%%%%%%%%
% APPENDIXES
%%%%%%%%%%%%%%%%%%%%%%%%%%%%%%%%%%%%%%%%%%%%%%%%%%%%%%%%%%%%%%%%%%%%%

%%%%%%%%%%%%%%%%%%%%%%%%%%%%%%%%%%%%%%%%%%%%%%
\appendix
\appendixtitle{Eulerian and Lagrangian analysis in the Gulf Stream region}\label{sec:app_gulfstream}

%%%%%%%%%%%%%%%%%
% FIG. A1
\begin{figure*}[t!]
    \centering
    \includegraphics[width=0.9\linewidth]{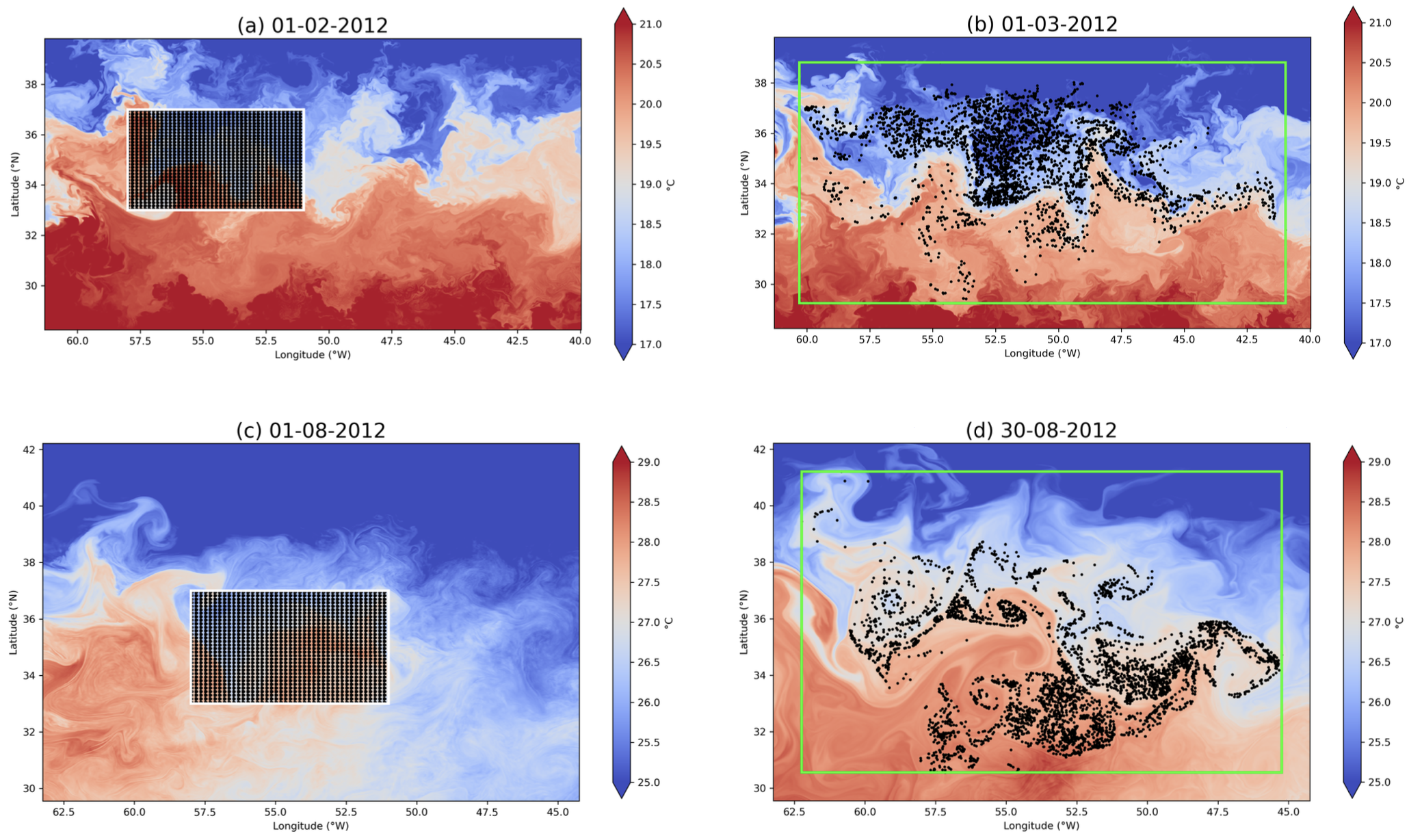}
    \caption{Snapshots of the SST field in the Gulf Stream region in February (top line)  and in August (bottom line) at the beginning (a, c) and at the end (b, d) of the $30$-day Lagrangian advection experiments. The corresponding particle distributions are shown with black dots. The green rectangles in (b, d) indicate the largest area covered by particles on the latest day of the month. 
    }
    \label{fig:fA1}
\end{figure*}
%%%%%%%%%%%%%%%%%

Here we present a more extensive characterization of the Eulerian flow properties and relative dispersion results in the Gulf Stream region, to contrast with those found in the Kuroshio Extension (see main text). 

For both winter and summer, the particle distributions at the beginning and at the end of the $1$-month advection period, superimposed over the simultaneous SST fields, are shown in Fig.~\ref{fig:fA1}. In February (Fig.~\ref{fig:fA1}a), the flow is characterized by a lot of mixed-layer instabilities, which reveal themselves in the roll-up of SST fronts at the smallest scales. 
In August, the spatial organization of the temperature field is driven by the presence of mesoscale features, such as large-scale filaments (Fig.~\ref{fig:fA1}c). 
The overall picture is analogous to the one in Kuroshio Extension.
Concerning the Lagrangian particle distribution, we see that, after one month, particles tend to spread more homogeneously in February, while they are more affected by the large-scale structures of the flow in summer (Figs.~\ref{fig:fA1}b,~d). 
These patterns suggest that dispersion is more local (i.e. more affected by smaller-scale flow features) in winter than in summer. 

%%%%%%%%%%%%%%%%%
% FIG. A2
\begin{figure*}
    \centering
    \includegraphics[width=0.9\linewidth]{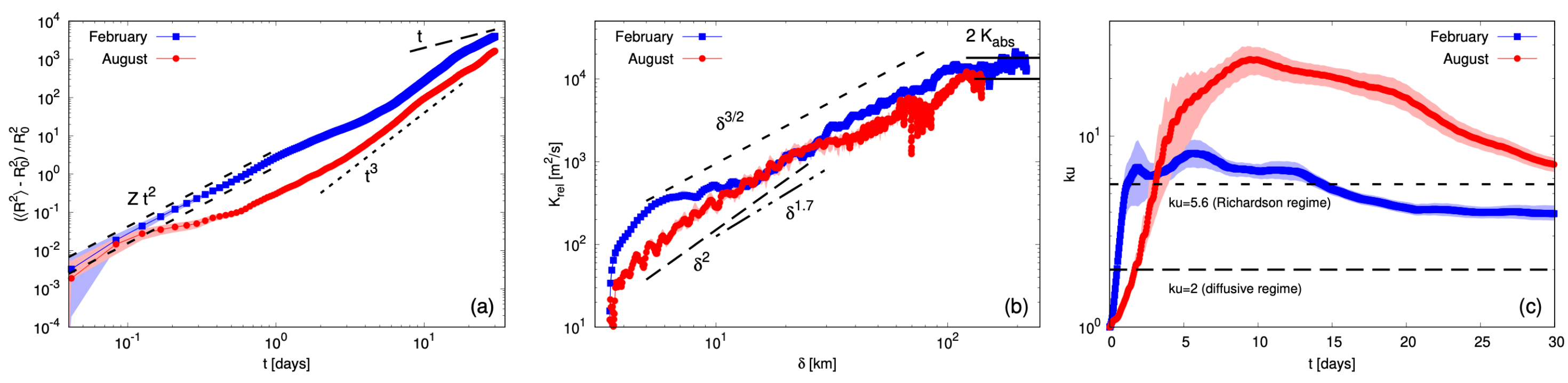}
    \caption{Normalized relative dispersion vs time (a), relative diffusivity vs the separation distance $\delta=\langle R^2(t)\rangle^{1/2}$ (b) and kurtosis vs time (c) in the Gulf Stream region, for February (blue squares) and August (red dots). Uncertainties are estimated as the $95\%$ confidence interval from a bootstrapping procedure. 
    } 
    \label{fig:fA2}
\end{figure*}
%%%%%%%%%%%%%%%%%
Figure~\ref{fig:fA2} shows complementary results from Lagrangian indicators completing those presented in Sec.~\ref{sec:llc_gs}, namely relative dispersion as a function of time, relative diffusivity versus the separation distance $\delta=\langle R^2(t)\rangle^{1/2}$ and kurtosis versus time. The general trends are quite similar to those found in Kuroshio Extension. Relative dispersion at short times here shows a clearer agreement with the prediction $Z t^2$ also in summer. In August, it later slows down [after $t \approx (0.1-0.2)$~days], before approaching a growth close to $t^3$ or slightly faster. 
In February, $\langle R^2(t) \rangle$ is generally larger at intermediate times. Its subsequent behavior is not very far from that of August (roughly $\sim t^3$), but less clear in terms of scaling. More generally, also in this region, it is not straightforward to identify dispersion regimes from this indicator. Relative diffusivity $K_{rel}$, when plotted against the separation distance $\delta = \langle R^2(t) \rangle^{1/2}$ more clearly allows to distinguish the winter and summer dispersion regimes. In February, $K_{rel}$ fluctuates around a $\sim \delta^{3/2}$ law, between $10$ and $100$~km, which would correspond to a spectrum $E(k) \sim k^{-2}$. Interestingly, however, in August we find a rather clear $\delta^2$ scaling from about $5$ to $50$~km, as one would expect for a spectrum steeper than $k^{-3}$ and pointing to nonlocal dispersion. Beyond this range, diffusivity shows slower growth, possibly suggestive of local dispersion, and roughly compatible with $K_{rel} \sim \delta^{3/2}$ (or the close scaling $K_{rel} \sim \delta^{1.7}$, corresponding to $\beta=2.4$, over a smaller subrange of separations). 
Correspondingly, while in winter kurtosis quite soon attains a constant value close to $5.6$ [the expectation for local, Richardson dispersion, for which $\langle R^2(t) \rangle \sim t^3$ and $K_{rel}(\delta) \sim \delta^{4/3}$] and stays close to it for almost all the advection period, in summer $ku$ initially grows to a value $5$ or $6$ times larger, before starting a slow decay after about $10$~days of advection. In the second half of August, these data do not allow to draw a safe conclusion on the dispersion regime, a longer simulation would be needed to clarify this point.

%%%%%%%%%%%%%%%%%%%%%%%%%%%%%%%%%%%%%%%%%%%%%%%%%%%%%%%%%%%%%%%%%%%%%
% REFERENCES
%%%%%%%%%%%%%%%%%%%%%%%%%%%%%%%%%%%%%%%%%%%%%%%%%%%%%%%%%%%%%%%%%%%%%
% Make your BibTeX bibliography by using these commands:
\bibliographystyle{ametsocV6}
\bibliography{references}

\end{document}